It Takes a Village:

Documenting the Contributions of Non-Scientific Staff to Scientific Research

Valerie Higgins

(ORCID: 0000-0001-7636-5039)

Fermi National Accelerator Laboratory

It Takes a Village:

Documenting the Contributions of Non-Scientific Staff to Scientific Research

Science, especially large-scale basic research, is a collaborative endeavor, often drawing on the skills of people from a wide variety of disciplines. These people include not just scientists, but also administrators, engineers, and many others. Fermilab, a Department of Energy National Laboratory and the United States' premier particle physics laboratory, exemplifies this kind of research; many of its high-energy physics experiments involve hundreds of collaborators from all over the world. The Fermilab Archives seeks to document the history of the lab and the unique scientific research its staff and visitors perform. Adequately documenting the lab's work often requires us to go far beyond things like the writings and correspondence of scientists to also capture the administrative and social histories of the experiments and the context in which they were performed. At Fermilab, we have sought to capture these elements of the lab's activities through an oral history program that focuses on support staff as well as physicists and collection development choices that recognize the importance of records documenting the cultural life of the lab. These materials are not merely supplementary, but rather essential documentation of the many types of labor that go into the planning and execution of an experiment or the construction of an accelerator and the context in which this work is performed. Any picture of these experiments and accelerators that did not include this type of information would be incomplete. While the importance and richness of this material is especially pronounced at Fermilab due to the massive size of its experiments and accelerator facilities and its vibrant cultural life, the fruitfulness of these collecting efforts at Fermilab suggests that other archives documenting modern STEM research should also make sure the contributions of non-technical and non-scientific staff are preserved and that researchers interested in this subject should not neglect such sources.

The United States Atomic Energy Commission established Fermilab in 1967 to construct what would become the world's most powerful particle accelerator. Originally known as the National Accelerator Laboratory, the vision for the lab was that it would develop, build, and operate accelerators that physicists from universities and other institutions all over the world would use to conduct high-energy physics experiments. In its early days, the lab's offices were located in a collection of houses called "The Village" that had once been a small town named Weston. While many of the lab's early employees were physicists and engineers involved in designing the accelerator, the 1969 Village telephone directory shows that even at this nascent stage, the lab depended on the contributions of other types of staff members. Areas like the Directorate, Beam Transfer, the Booster, Experimental Facilities, the Linac, the Main Accelerator, and Theory each have one or more administrative assistants, and other members of the staff include technicians, drivers, a librarian, an artist, a photographer, a nurse, an activities coordinator, accountants, purchasing administrators, warehouse workers, maintenance men, construction inspectors, switchboard operators, a food service manager, a director of public information, and many other positions someone unfamiliar with the lab might not expect.

From early on in the lab's history, many of the people involved recognized the historical significance of the lab and its work. The Fermilab History and Archives Office traces its origins back to the Fermilab History Committee, which was established by first lab director Robert R. Wilson in January 1977, as the lab approached its tenth anniversary. The lab's first archivist, Lillian Hoddeson, began collecting Fermilab historical records and artifacts and conducting oral history interviews in January 1978. The mission of the History and Archives Office is to identify, collect, organize, preserve, and make accessible records of enduring value that document the activities of Fermi National Accelerator Laboratory and the field of high-energy physics. To accomplish these goals, the Fermilab Archives acquires records (including both print and digital) and artifacts created by Fermilab as well as other organizations and individuals. These records include a broad range of materials, such as correspondence

4
DOCUMENTING NON-SCIENTIFIC STAFF IN SCIENCEfiles maintained by the Directorate, logbooks for accelerators and experiments, clippings of media articles on the lab collected by the Office of Public Information, original artwork by a longtime lab artist, and lab and experiment websites, to name just a few. From the earliest days of the Archives, the lab archivist has also recorded oral history interviews that complement these collections. The work of the Archives supports the activities of Fermilab's staff, its users, and outside researchers interested in the history of Fermilab or of particle physics and related disciplines. Some of the Archive's most frequently used materials are stored in an office called the Milton G. White History of Accelerators Room on the third floor of Fermilab's central office building, Wilson Hall. The majority of the collections are stored in one of the original Weston houses in the Fermilab Village. This house was renovated in the early 1990s to make it suitable for archival storage. Due to the limited space in the Wilson Hall Office and the Archives House, a significant portion of the collections are also stored in an offsite records storage facility. In total, the Archives holds about 2,000 linear feet of material, plus additional oversized materials and about two terabytes of digital materials.

Today, Fermilab continues to build world-class particle accelerators and hosts high-energy physics experiments, many of which are performed by large, international collaborations. Building an accelerator or performing an experiment can take many years, even decades, and these efforts require the work not only of scientists and engineers, but also computer programmers, technicians, and others, including non-technical staff. The majority of Fermilab's employees are non-scientists, and it is essential for the Fermilab Archives to document the contributions of these people to get a complete picture of the work the lab does.

Technical but non-scientific staff are one of the most obvious groups whose contributions should be documented by any archive working with large-scale scientific research. Fermilab's experiments and accelerators push at the edges of human knowledge, so they often require lab staff to build unique engineering and computing solutions. Documenting these efforts is just as crucial as documenting the work of scientists to create and preserve a full picture of the history of the lab's projects. For example, the collections of the Fermilab Archives include the papers of a programmer who played a key role in the development of Fermilab's early data-taking software. We also collect the records of high-level computing administrators, but the records of programmers who worked "on the ground" with physicists conducting significant experiments can show how they interacted with the physicists to produce the specialized software needed to collect and analyze the data from those experiments. Some of these programmers are trained as physicists themselves, so the division between them and the scientists is not always as clear as titles might suggest. Oral history interviews with longtime employees in Fermilab's core and scientific computing divisions are also a valuable source of information about how the lab's computing professionals interacted with other areas of the lab and how the role of computing has changed over the lab's history. As in many other fields, the role played by computing in high-energy physics research has steadily increased, making these records and interviews increasingly valuable.

While the importance of documenting the contributions of technical staff like computing professionals to scientific research might be relatively obvious, the contributions of generally non-technical staff are also an essential part of the story of scientific research. It is not always easy to define which roles should be considered technical and which non-technical; the lab's first director, Robert Wilson, felt that physicists should run as many areas of the lab as possible, including functions like personnel and the business office (Hoddeson, Kolb, & Westfall, 2008, p. 110), and many ostensibly non-technical positions required workers to learn highly specialized skills. When examining an organizational chart or similar documentation of a scientific project, a person's title and position in the organization alone may not tell you what knowledge they have or what ways in which they might have contributed to the work; often, people who contributed beyond what their titles might indicate can only be identified by speaking with others whose key roles in the function or project to be documented are more obvious.

Long-time administrative assistants in the Directorate and scientific areas at the lab are a particularly valuable source of information about the lab's operations. They are often involved in many projects over the course of their careers and are familiar with the administrative history of those projects. This kind of information is especially valuable when trying to understand the early operations of Fermilab; the lab's first director so disliked hierarchies and bureaucracy that the lab never had a stable enough organization chart to send to the Atomic Energy Commission in the entire eleven years he led the lab (Hoddeson, Kolb, & Westfall, 2008, p. 110). In some cases, an administrative assistant might serve in a department longer than any of its leader and can be knowledgeable about ways in which the operations of the department changed over time. They can often shed light on questions like what pressures the departments they worked in faced, who spearheaded the creation of certain projects, and what needs led to the creation of new positions. Even in organizations with more rigid organizations than Fermilab had in its first decade, administrative assistants are likely to still be good sources of information about the operations of the areas in which they worked.

As an organization becomes more complex and sprawling, this sort of information about how its organizational structure developed and to what extent that structure reflects the actual operations of the organization becomes increasingly valuable. Modern scientific research is often conducted on an enormous scale, with projects that last for many years and involve hundreds, even thousands of people. The growth of experiments that Fermilab has been involved in reflects this. When Fermilab experiment E-288 discovered the bottom quark in 1977, that collaboration included sixteen people. When Fermilab's experiments CDF and DZero announced the discovery of the top quark in 1995, each collaboration included about 450 people from institutions all over the world. When the experiments CMS, in which many Fermilab scientists participate, and ATLAS at CERN's Large Hadron Collider discovered the Higgs boson in 2012, those collaborations each had about 3,000 participants. Projects like these become long-lasting institutions unto themselves with their own organizational structures that may spread across several institutions. This kind of growth indicates that it will be increasingly important to document the contributions of the administrative people who help keep those large projects running. Their knowledge is essential to developing a full picture of how these large-scale projects function.

People in these administrative assistant positions at Fermilab often develop specialized skills. One long-time administrative assistant states in an oral history interview that she has the "highest regard for secretaries" who worked at the lab because they were often called on to perform a wide range of duties and learned to do things like type up papers with complex equations for the physicists. Another administrative assistant noted that, after years of experience, she could sometimes spot errors in the papers because something would look different than what she had typed before, even if she didn't understand the content. These employees often played leading roles in the organization of conferences at which physicists shared their experimental results and planned the future directions of the lab, as well. This illustrates that these were skilled positions and that the people who held them were deeply involved in many aspects of the lab's operation.

Additionally, administrative assistants can often bring unique perspectives to the archival record. As Fermilab's early loose structure illustrates, organizational charts can only tell an archivist or researcher so much about how a lab or experiment actually operated. Administrative assistants are repositories of institutional knowledge that is often not captured in records. Fermilab administrative assistants frequently interact with large numbers of people from a wide range of areas at the lab and experiment collaborators from all over the world, so they know how different areas of the lab work together and know what sort of informal lines of communication exist that might not be reflected in reporting structures. They usually become very familiar with the scientists and administrators with whom they work, so they often know how people in leadership positions made decisions. For instance, Wilson's first principal administrative assistant had previously worked as an administrative assistant to Ernest Lawrence and Robert Oppenheimer at Los Alamos, and in an oral history interview shares her observations about their different leadership and decision-making styles. According to one long-time administrative assistant who served under multiple lab directors, the directors even turned to her for advice on some decisions that involved organizing people because they knew she had valuable knowledge about the personalities of people involved and had seen how such things had been handled in the past.

Administrative assistants and other non-technical staff can also provide unique insights into the culture of their workplaces. Like all other human activities, any scientific research project is conducted within a particular social context, and it is important to understand that context in order to fully understand the research and the people who produced it. Awareness of the ways in which the physical and social environment impacts the conduct of science has long been part of Fermilab's ethos. When Wilson accepted his position as Fermilab's first director, he set out to create a site that was physically beautiful in order to draw the world's best high-energy physicists to the new lab and to inspire creativity and productivity among the lab's staff. He and his deputy director also set out to create an inclusive work environment, stating in their 1968 "Policy Statement on Human Rights,"

> It will be the policy of the National Accelerator Laboratory to seek the achievement of its scientific goals within a framework of equal employment opportunity and of a deep dedication to the fundamental tenets of human rights and dignity… Prejudice has no place in the pursuit of knowledge… It is essential that the Laboratory provide an environment in which both its staff and its visitors can live and work with pride and dignity.

While scientists and engineers can, of course, also shed light on the social contexts in which scientific work is performed, non-technical staff often have particularly insightful observations to share. People like administrative assistants are deeply embedded in the workplace culture, but are also somewhat outside the professional culture of the physicists and often bring useful perspectives when they reflect on those cultures. Even today, the field of high-energy physics is still largely dominated by men, and one administrative assistant shared interesting observations about some of the challenges she saw past women scientists and administrators encounter in trying to get their ideas heard in meetings.

Another category of non-technical staff who can provide valuable insights into the workplace cultures in which science is conducted are members of departments that perform human resource functions. At Fermilab, interviews with people who served in the Workforce Resource and Development Section and its predecessors for many years have provided excellent insights into the lab's workplace cultures and ways in which they have changed over the years. Fermilab also has an unusually active cultural life, with a public arts and lectures series, an art gallery, numerous clubs, including some clubs for scientists from particular nations, and more, and these amenities have been a significant part of many staff members' experiences. The documentation of these activities, in which non-technical staff often played a significant role, is essential to documenting the history of the lab's culture and how it impacted the ways in which the lab's employees, including scientists and engineers, approached their work.

In general, this kind of valuable information about workplace culture, how decisions were made, ways in which members of organizations operated outside their organizational structures, the personalities of people in leadership roles, and more will not be found in the personal records of administrative assistants and other non-technical staff; the best way to capture it is through oral history interviews. These interviews make the memories and observations of non-technical staff part of the archival record. The Fermilab Archives has had an oral history program since 1978, and one of the earliest interviews the lab's archivist conducted was with Wilson's first administrative assistant, showing that there was some recognition even in the early days of the lab that people in these types of positions could have useful information about the history of the lab. Another interviewee from 1978 was the man who was the first director's driver. However, the early focus of the Archives' oral history program was clearly on scientists, and only a handful of the earliest interviews are with non-scientists. As time went on, the Archives expanded its oral history program to include more people in different roles at the lab. In addition to numerous administrative assistants, our collections include interviews with the lab site manager, a nurse from the lab's cancer therapy facility, the head of the lab's Equal Employment Opportunity Office, a lab photographer, a physicist's wife who played a key role in starting some of the lab's cultural activities, a lab librarian, and others. In recent years, we have begun interviewing many more non-technical staff members. In the past year, about half of my oral history interviews have been with non-scientists.

Generally, we conduct interviews individually, but group interviews may also be a valuable source of information. We have one example of a group interview that the lab archivist and historian conducted in 2003 with five long-serving administrative assistants at once. While this format does not allow any one person to go into great depth about their experiences, they remind one another of people and events, often prompting each other to recall stories they had forgotten. This is particularly valuable when trying to reconstruct information that is not preserved in records. I am interested in exploring this format more in the future as a supplement to individual interviews.

The importance of documenting the contributions of non-technical staff is illustrated by what is lost when those contributions are not documented. One of Fermilab's early particle detectors was the 15-foot bubble chamber, which holds a superheated liquid. When a moving charged particle entered the liquid, it would cause the liquid to boil along its path, creating a trail of minute bubbles. Different types of particles left distinctly different trails. Cameras in the bubble chamber took many thousands of photographs of these processes, and people known as scanners examined the images to identify the types of particle tracks the physicists instructed them to find. These scanners, who were mostly women, often had little or no technical background and learned the unique skills they needed on the job. So far, I have not found any oral history interviews with these scanners in our collections, and the records we have that relate to their work do not capture their personal experiences and impressions. Their work provided essential support to the work the scientists were doing, and having this type of information would provide a fuller picture of how these experiments achieved their results.

It is important to note that this emphasis on the value of oral history interviews for documenting the contributions of non-technical staff to the lab's operations does not mean that paper and digital records related to their contributions should be neglected. While interviews are often the best source of information on how they contributed to the lab's scientific projects and the best documentation of their unique perspectives, the records they produce can also be helpful for understanding their role in scientific research. The mission of the Fermilab Archives is to document the whole of the lab, including non-technical areas, so these records would fall under our collecting policy in any case. Even if our mission was only to document the history of the lab's scientific work, however, collecting records from administrative assistants, computer programmers, and areas of the lab like the Education Office, the Art Gallery, and other locations would be essential to understanding the scientific work the lab does.

It is also important to remember the limitations of oral history interviews. Memories are often imperfect, imprecise, and, in some cases, even inaccurate. Oral histories should be used in a way that acknowledges their limitations while drawing on their strengths; for instance, an oral history interview may not be the best resource from which to construct a detailed timeline of an experiment, but it might be the best way to learn about a lab director's management style. The limitations of memory are, in fact, one of the reasons an archive should have as broad an oral history program as possible—interviewing many different people from a wide range of positions gives researchers the ability to compare multiple recollections of the same events or subjects to find places in which those recollections agree and disagree. A robust archival program can also address some of the potential weaknesses of oral histories by ensuring that researchers have access to records that complement the interviews. This documentation can provide broader context for the stories shared by the interviewees, and researchers can check statements made by the interviewees that are verifiable, such as dates and position titles. Researchers and archivists should remember, however, that records may also contain errors or may not capture ways in which day-to-day realities diverged from what was documented in the records. Additionally, as many of the previous examples illustrate, whatever the flaws of oral histories, they are often the only source of information about things not captured in records.

In many ways, Fermilab is a unique institution. It has a rich cultural life, high-energy physics collaborations are very large, even for modern scientific research, and it had a very loose organizational structure in its early days. These and other qualities make the kinds of records and interviews I've described particularly essential for documenting Fermilab's work. However, the usefulness of these materials at Fermilab suggests that other archives and researchers interested in the history of modern scientific research should also seek out these types of materials. They are essential for creating a complete picture of all the types of labor that go into conducting scientific research, the ways in which people conducting scientific research organize themselves and communicate with one another, and the ways in which the context in which that research is performed influences how it is conducted and influences the people conducting it.